\documentclass[aps,prd,showpacs,twocolumn,floatfix]{revtex4}
\usepackage{graphicx}
\usepackage{amsfonts}
\usepackage{amssymb}
\usepackage{amsmath}
\usepackage{flafter}
\usepackage{epstopdf}
\usepackage{hyperref}
\usepackage{xcolor}
\usepackage{float}
\usepackage[stable]{footmisc}

\begin{document}

\title{Did Gamma Ray Burst Induce Cambrian Explosion?}

\date{\today} 
\author{Pisin Chen}
\affiliation{Department of Physics and Graduate Institute of Astrophysics, National Taiwan University, Taipei, Taiwan 10617}
\email{pisinchen@phys.ntu.edu.tw}
\affiliation{Leung Center for Cosmology and Particle Astrophysics (LeCosPA), National Taiwan University, Taipei, Taiwan, 10617}
\affiliation{Kavli Institute for Particle Astrophysics and Cosmology, SLAC National Accelerator Laboratory, Menlo Park, CA 94025, U.S.A.}

\author{Remo Ruffini}
\affiliation{Department of Physics, University of Rome, La Sapienza, Rome, Italy}
\affiliation{International Center for Relativistic Astrophysics Network, Pescara, Italy}
\email{ruffini@icra.it}

\addtocounter{footnote}{-20}

\begin{abstract}

One longstanding mystery in bio-evolution since Darwin's time is the origin of the Cambrian explosion that happened around 540 million years ago (Mya), where an extremely rapid increase of species occurred. Here we suggest that a nearby GRB event ~500 parsecs away, which should occur about once per 5 Gy, might have triggered the Cambrian explosion. Due to a relatively lower cross section and the conservation of photon number in Compton scattering, a substantial fraction of the GRB photons can reach the sea level and would induce DNA mutations in organisms protected by a shallow layer of water or soil, thus expediting the bio-diversification. This possibility of inducing genetic mutations is unique among all candidate sources for major incidents in the history of bio-evolution. A possible evidence would be the anomalous abundance of certain nuclear isotopes with long half-lives transmuted by the GRB photons in geological records from the Cambrian period. Our notion also imposes constraints on the evolution of exoplanet organisms and the migration of panspermia.

\vspace{3mm}


\end{abstract}

\maketitle


There exists a unique ``mass genesis" in Earth's history, that is the Cambrian explosion, the so-called ``bio-evolution's big bang", where diversification of life accentuated from around 540 to 520 Mya \cite{Bowring}. All major phila were established during that time (see Fig. 1). The origin of this Cambrian explosion remains a mystery ever since the mid 19th century. Charles Darwin in 1859 discussed about it in his famous treatise {\it On the Origin of Species by Natural Selection} as ``one of the main objections that could be made against the theory of evolution by natural selection" \cite{Darwin}. 
``The controversy has changed character in the 150 years since this problem was first posed. But  the central question, which remains unresolved, is: why did early animals wait until the Cambrian boundary to make their appearance?" \cite{McMenamin}.

\begin{figure}
\includegraphics[width=8.5cm]{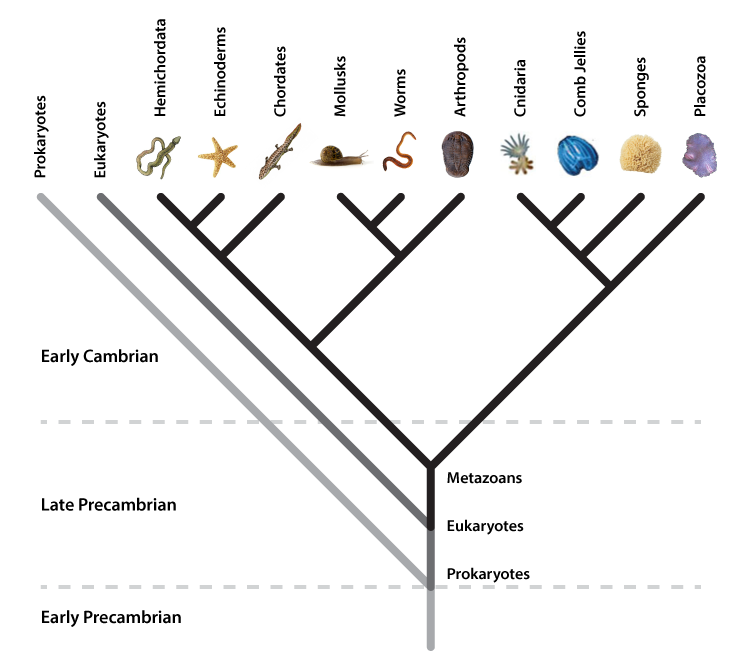}
\caption{Figure 1. Cambrian Explosion: Evidence in fossil records show that all major phyla were established around 540 Mya in the Cambrian period.}
\end{figure}

Since the Cambrian explosion, there had been five mass extinctions \cite{MassExtinction}. Extraterrestrial impacts such as asteroids, supernovae, and gamma ray bursts (GRB) have been proposed as the sources for some of these events \cite{Alvarez,Melott2004}. 
Here we suggest that the Cambrian explosion might have been triggered by a nearby GRB event. Due to a relatively lower cross section, but more importantly the conservation of photon number in Compton scattering, a fraction of the GRB photons can penetrate the entire atmosphere to reach the sea level, albeit with a modified spectrum. Organisms under shelter or protected by a shallow layer of soil or water would possibly survive but suffer from induced DNA mutations, which resulted in accelerated bio-diversification. 

We are interested in `nearby' GRB events within our Milky Way galaxy in a relatively recent time, e.g., $<$ 5 Gy. This means that our estimate of the GRB event rate should rely on the short bursts, whose estimate varies. One estimate gives one event per 0.1-1 My in Milky Way galaxy \cite{Lattimer}. Coward et al. on the other hand estimated, based on the {\it Swift} satellite telescope data, a rate density of 8-1100 ${\rm Gpc}^{-3}{\rm y}^{-1}$ \cite{Coward}. Assuming 1 Mpc  for the average distance between galaxies, this translates into a rate of one event per 1-5 My per galaxy. We shall take the reference value 1 per 5${\rm  My}^{-1}$ per galaxy in our discussion. As a rough estimate, we model Milky Way galaxy as a two-dimensional disk with a radius of 15 kpc. Then the distance for a GRB to occur once in Earth's entire history, for example, is $d\sim 500$ pc ($\sim$ 1500 lightyears) from the solar system.

Evidences show that the Cambrian atmosphere was predominantly made of nitrogen with a density that is comparable to the present level, while the abundance of oxygen was only a few percent of the present value \cite{Borzenkova}. So in our estimate we model the Cambrian atmosphere as 
\begin{equation}
n(h)=n_0\Big(1-\frac{h}{H}\Big)^{k-1}, \quad\quad h \leq H,
\end{equation}
where $n_0\sim 5\times 10^{19}$ ${\rm cm}^{-3}$ is the present ${\rm N}_2$ number density at the sea level, $H\sim 44$ ${\rm km}$ the total thickness of the atmosphere, and $k\sim 5.5$.

The photon-atom interaction cross section is dominated by three physical processes, namely the photoionization, the Compton scattering, and the $e^+e^-$ pair production at low, medium and high energies, respectively \cite{Samson}. 
For nitrogen atoms, the photoionization dominates for photon energies below $\sim 50$ keV with the cross section \cite{Samson}
$\sigma_i(x)=3.67\times 10^4 (xm_e[{\rm eV}])^{-2.3}$ Mb, where $x\equiv \nu/m_e$, and $m_e\approx 500$ keV is the electron rest mass. 
It reduces drastically from $\sim 183$ Mb at $x \sim 2\times 10^{-5}$ to $\sim 4.5$ b at $x \sim 0.1$. In the range $\sim 50 {\rm keV}-50 {\rm MeV}$, Compton scattering dominates, with the cross section per nitrogen atom roughly equals to $Z=7$ times the Klein-Nishina formula under the free-electron approximation:  
\begin{equation}\label{ComptonScattering}
\frac{d\sigma_c(x)}{d\Omega} = \frac{7 r_e^2}{2}\Big(\frac{x'}{x}\Big)^2\Big(\frac{x}{x'}+\frac{x'}{x}-\sin^2\theta\Big),
\end{equation}
where $x$ and $x'$ are the initial and final photon energy, respectively, and $\theta$ is the angle between the two, which satisfy the kinematic relation: $x-x'=xx'(1-\cos\theta)$. The total cross section is obtained by integrating over the solid angle. It decreases mildly from 3.9 b at $x=0.1$ to 0.1 b at $x=100$. 
Finally at energies above $\sim$50 MeV, the photon energy loss is dominated by the pair production process, which is rather insensitive to photon energy and is at the sub-barn level. 

As we will see below, the GRB spectrum diminishes at high energies. On the other hand, the portion dominated by photoionization, whose cross section is large, cannot survive to the sea level. Thus only the portion ($x\sim 0.1-100$) dominated by Compton scattering may contribute effectively to the final flux at sea level. One salient character of Compton scattering is that the process preserves the photon number. In addition, a good fraction of the outgoing photon moves in the forward direction, which may survive to the sea level cascading through successive Compton scatterings. We will make such estimate below. 

The GRB spectrum can be described by the Band function \cite{Band} (with $\alpha=-1, \beta=-2$, for simplicity):
\begin{equation}\label{BandFunction2}
N(x)= N_0
\begin{cases}
e^{-x/x_0}(1/x),  &x\leq x_0, \\
e^{-1}(x_0/x^2),  &x>x_0,
\end{cases}
\end{equation}
where $x_0$ is the break of the spectrum, typically $x_0\sim {\mathcal O}(1)$, i.e., a few hundred keV. 
We follow Melott et al. \cite{Melott2004} to assume that the {\it isotropic equivalent energy} released by a GRB is $\sim 5\times 10^{44}\mathrm{W}$. Then the flux of a GRB event at distance $d$ is 
$F_{\nu}(d)=m_e^2\int xN(x)dx/4\pi d^2 \simeq 5\times 10^{51}/4\pi d^2$ $\mathrm{[erg/cm^2/sec]}$.
For example, $F_{\nu}(d=500 {\rm pc})\sim 2\times 10^{8}$ erg/cm$^2$/sec for $x_0=1$, with a corresponding $N_0\approx 5 \times 10^{19} {\rm /erg/cm}^2/{\rm sec}$. 
The prompt signals of a short-burst typically lasts for 1-2 sec. So the total flux is roughly the same amount.

To estimate the final flux of GRB at the sea level, we employ a forward-cone approximation, where outgoing photon with angle $\theta > \theta_c=1$ is considered lost while inside this forward-cone an averaged out-going energy, $\langle x' \rangle= [\int^{\theta_c}_{0}x'(\theta)(d\sigma_c/d\Omega)d\cos\theta]/[\int^{\theta_c}_{0}(d\sigma_c/d\Omega)d\cos\theta]$, is invoked in tracking the evolution of the Compton cascade. The result is plotted in Fig. 2. The integrated GRB final flux is reduced from that of the initial by a factor $\sim 5\times 10^{-5}$. 

\begin{figure} 
\includegraphics[width=8.5cm]{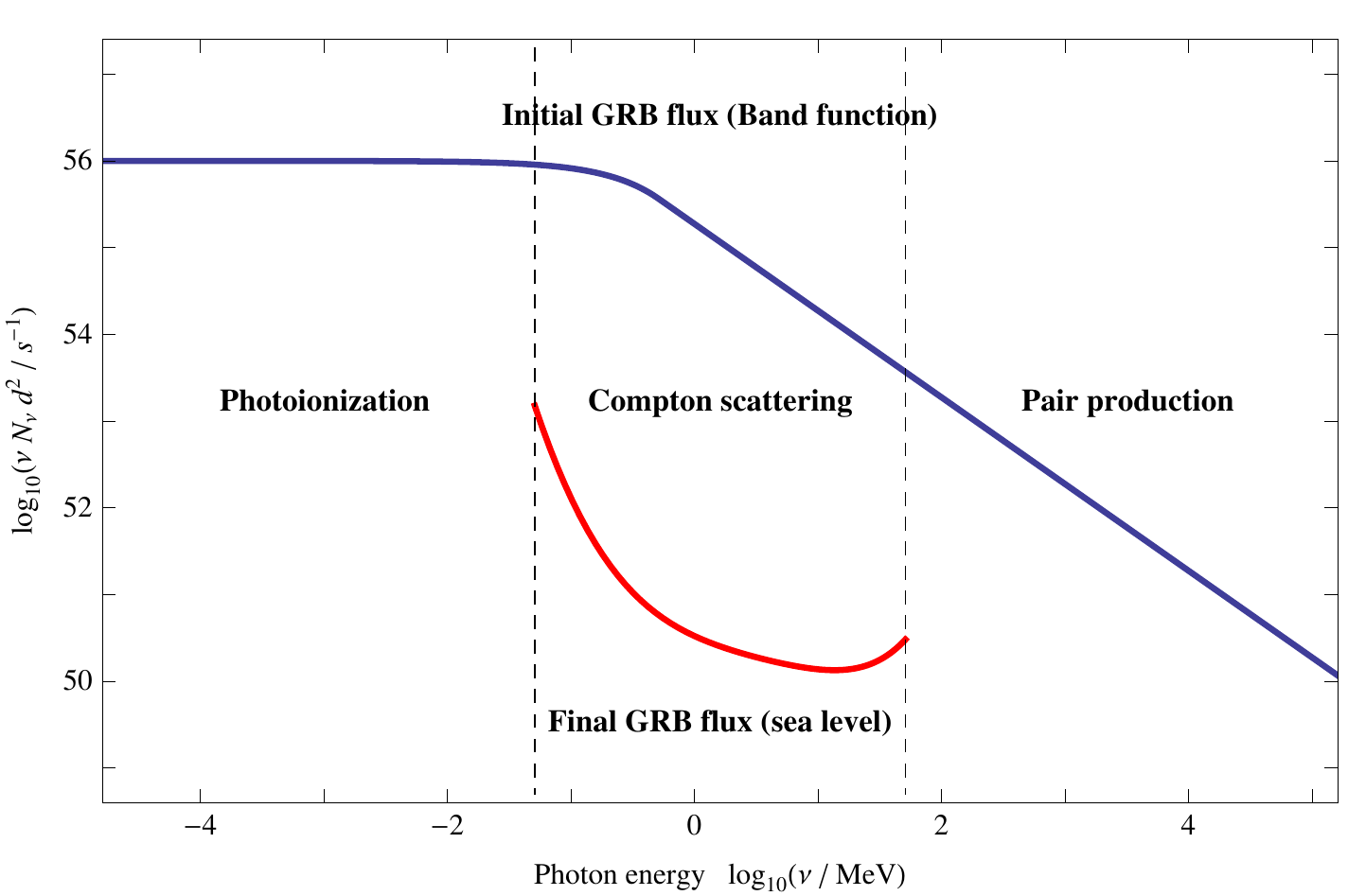}
\caption{Figure 2. Normalized Initial (Band function) vs. final (sea level) GRB flux as a function of photon energy range (50 keV to 50 GeV). The latter is made under a forward-cone approximation for Compton scattering, while the photons in the sections of the spectrum dominated by photoionization (left) and pair production (right) are assumed to be totally lost.}
\end{figure}

It is known that over-dosage of x-ray would induce DNA mutations \cite{Russell,McClean}. Experiments show that such induced mutations is much more effective than the spontaneous ones \cite{Griffith}. However, the effect is complex and thus difficult to be characterized quantitatively and generically. 
To put the issue in perspective, however, we may take the annual radiation dosage limit for a human body, $\sim 5$ ${\rm rem}$ (1 ${\rm rem}\equiv$ 1 ${\rm erg/g}$, or 1 ${\rm erg/cm}^3$ for water) \cite{AmNucSoc}, as a reference. Such dosage limit should be commensurate with the number of cells of an organism. Primordial marine organisms are much smaller in size than modern humans, so the dosage limit should in principle be much lower than that for the human. The amount of radiation dosage estimated above would be fatal for organisms exposed in air. Luckily, most organisms in the Cambrian time were in shallow waters \cite{Bowring,McMenamin}. 

The mean-free-path of hard x-ray photons in water is $l\sim \mathcal{O}(10)$ ${\rm cm}$. Interacting with water molecules, these hard x-rays would soon cascade to softer photons and electron-positrons and deposit their energies to water and marine organisms. For a GRB event occurring at a distance $d=500$ ${\rm pc}$, the energy deposition to the ocean surface water is roughly $\sim \mathcal{O}(10^3)$ ${\rm rem}$ in the first mean-free-path. Marine organisms living several mean-free-paths below surface might survive the impact but suffer induced DNA mutations. 

Unlike the `big five' mass extinctions, Cambrian explosion is unique in that it apparently occurred without a companion mass extinction. GRB is the only one among all proposed terrestrial and extraterrestrial sources for mass extinctions that can provide an explanation to this mass genesis. Without an obvious precursor mass extinction, the Cambrian explosion cannot be explained by the decimation of the eco-system. But why did it wait for 3-4 Gy to happen? GRB occurred stochastically. As we have shown above, a nearby event that happened once every 5 Gy appears to provide the right amount of radiation for triggering induced DNA mutations for organisms in water. More distant, thus more frequent events, say that happened once per 100 My, would have a final flux at the sea level that is $\sim \mathcal{O}(100)$ times weaker, which may not be as effective in triggering mass genesis.   

If the same GRB event occurred in a different era that had a much denser atmosphere, then it would have much less impact on creatures. On the contrary, if the air was much thinner than today's, then it might have wiped out lives entirely. This may have implications to the extinction of life on Mars, whose atmosphere is much thinner.

One possible evidence for this proposed origin of Cambrian explosion would be the anomalous abundance of certain isotopes in geological records from the Cambrian period. In the ``giant resonance" region of photonuclear reactions, which corresponds to photon energy $\sim15-30$ MeV, the cross sections are large and nuclear transmutation can be induced effectively \cite{Ejiri}. Transmuted isotopes with half-lives $\sim \mathcal{O}(100)$ My or longer may then leave imprints of this GRB incident. 

Our notion also imposes constraints on exoplanet life and panspermia. Exoplanets within 500 pc from our solar system would subject to a similar stimulation of DNA mutations that would accelerate its bio-evolution if it has an atmosphere similar to that of the Earth. Unprotected primitive microscopic organisms carried by interstellar rocks may turn sterile after been exposed to GRB at certain distance. However, such seeds of panspermia would evade GRB destruction if its migration and colonization rate is faster than the GRB event rate within a volume of destruction. 

{\bf Acknowledgement}
We thank Ronald Adler and Lance Labun for reading the manuscript and providing useful comments. P.C. also gratefully acknowledges helpful discussions with Chia-Wei Lee during the initial stage of his investigation and the technical assistance of Nicholas Yi-Huan Chen for preparing Figure 1 and Yu-Hsiang Lin for calculating the flux and plotting Figure 2. P.C. is supported by Taiwan's National Science Council and by the Leung Center for Cosmology and Particle Astrophysics, National Taiwan University.


\begin{references}

\bibitem{Bowring}
Bowring, S. A. et al., ``Calibrating rates of early Cambrian evolution". Science {\bf 261}, 1293 (1993).

\bibitem{McMenamin}
McMenamin, M. A. S. and McMenamin, D. L. S., {\it The Emergence of Animals: The Cambrian Breakthrough}, Columbia University Press (1990).

\bibitem{MassExtinction}
Hallam, A. and Wignall, P. B., {\it Mass Extinction and Their Aftermath}, Oxford University Press (1997).

\bibitem{Darwin}
Darwin, C., {\it On the Origin of Species by Natural Selection}, pp. 306Ð308, London: Murray (1859).

\bibitem{Alvarez}
Alvarez, L. W., Alvarez, W., Asaro, F., Michel, H. V., ``Extraterrestrial cause for the Cretaceous-Tertiary extinction". Science {\bf 208}, 1095 (1980).

\bibitem{Melott2004}
Melott, A. L. et al., ``Did a Gamma-Ray Burst Initiate the Late Ordovician Mass Extinction?". Int. J. Astrobiology {\bf 3 (1)}, 55 (2004).

\bibitem{Lattimer}
Lattimer, J. M., ``Lecture notes on gamma ray bursts".
http://www.astro.sunysb.edu/lattimer/AST301/lecture-grb.pdf

\bibitem{Coward}
Coward, M. D. et al., ``The Swift short gamma-ray burst rate density: implications
for binary neutron star merger rates". Mon. Not. R. Astron. Soc. {\bf 425}, 2668 (2012) [arXiv:1202.2179v2]. 

\bibitem{Borzenkova}
Borzenkova, I. I. and Turchinovich, I. Yu.,  ``History of atmospheric composition", in {\it Environmental Structure and Function: Climate System}, Vol. II, ed. G. V. Gruza, EOLSS, UNESCO (2003).

\bibitem{Samson}
Samson, J.A.R. and Angel, G. C., ``Single- and double-photoionization cross sections of atomic nitrogen from threshold to 31 \AA". Phys. Rev. A {\bf 42}, 1307 (1990).

\bibitem{Band}
Band, D. et al., ``BATSE observations of gamma-ray burst spectra. I: Spectral diversity". Astrophys. J. {\bf 413}, 281 (1993).


\bibitem{Russell}
Russell, W. L., ``X-ray-induced mutations in mice", Cold Spring Harb. Symp. Quant. Biol. {\bf 16}, 327 (1951).

\bibitem{McClean}
McClean, P., {\it Genes and Mutations},

[http://www.ndsu.edu/pubweb/~mcclean/] (1999).

\bibitem{Griffith}
Griffith, A. J. F., Miller, J. H., Suzuki, D. T. et al., {\it An Introduction to Genetic Analysis}, 7th edition,  W. H. Freeman, New York (2000).

\bibitem{AmNucSoc}
See, for example, the American Nuclear Society homepage:
http://www.ans.org/pi/resources/dosechart/.

\bibitem{Ejiri}
Ejiri, H, et al., ``Resonant Photonuclear Reactions for Isotope Transmutation", J. Phys. Soc. Japan {\bf 80}, 094202 (2011).

\end{references}
\end{document}